\documentclass[12pt]{article}

\usepackage{amsmath}
\usepackage{amscd}
\usepackage{amssymb}
\usepackage{amsfonts}
\usepackage[mathscr]{euscript}
\usepackage{pstricks}
\usepackage{pst-node}
\usepackage[dvips]{pstcol}

\newcommand{\Tr}{\operatorname{Tr}}

\title{\bf Complete positivity and dissipative factorized dynamics}

\author{Fabio Benatti$^{a,b}$, Roberto Floreanini$^{b}$, 
Raffaele Romano$^{a,b}$\\
\small${}^a$Dipartimento di Fisica Teorica, Universit\`a di Trieste,
Strada Costiera 11,\\
\small 34014 Trieste, Italy\\
\small ${}^b$Istituto Nazionale di Fisica Nucleare, Sezione di Trieste,
34100 Trieste, Italy} 

\date{\null}

\begin{document}

\maketitle

\vskip 2cm

\begin{abstract}
\noindent
We show that any Hermiticity  and trace preserving continuous
semigroup $\{\gamma_t\}_{t\geqslant 0}$ in $d$ dimensions is
completely positive if and only if the semigroup $\{\gamma_t \otimes
\gamma_t\}_{t\geqslant 0}$ is positivity preserving.
\end{abstract}

\vfill\eject

%%%%%%%%%%%%%%%%%%%%%%%%%%%%%%%%%%%%%%%%%%%%%%%%%%%%%%%%%%%%%%%%%%%%%%%%%%%%%%

Complete positivity is a property of linear transformations of quantum
states whose importance for physics is prominent in open quantum
system dynamics and quantum communication.

The reduced dynamics of systems in (weak) interaction with
their environment is usually generated by equation of motion of
Kossakowski-Lindblad form~\cite{spoh} 
and thus completely positive~\cite{gori,lind}. 
However, it is
still being debated whether such a constraint is physically
necessary~\cite{silb,pech,haak,stru,gasp}.

On the contrary, in quantum communication theory only completely
positive linear maps can describe local operations 
on quantum states~\cite{niel}.
Locality means that, given a bipartite system $A + B$ in a state
$\rho_{AB}$, only the $A$-component is trasformed according to
$\gamma_A\otimes{\bf I}_B$, where ${\bf I}$ is the identity operation; 
then, if $\rho_{AB}$ is entangled and
$\gamma_A$ not completely positive, $\gamma_A \otimes{\bf
I}_B[\rho_{AB}]$ may develop negative eigenvalues and thus lose
consistency as a physical state~\cite{krau,horo,ben1}.

The same kind of argument is generally used to motivate why the
reduced dynamics of an open quantum system $A$ must be described by a
semigroup of completely positive dynamical maps $\gamma_t^A$; if not,
$(\gamma_t^A \otimes{\bf I}_B)[\rho_{AB}]$ may become physically
inconsistent as the time evolution of an initial entangled state
$\rho_{AB}$~\cite{spoh}. 

In this case, however, the partner system $B$ is not, as
in quantum communication theory, a concrete party, making up, for
instance, a definite protocol for information transmission. 
Rather, $B$ is a totally uncontrollable entity that may happen to have became
entangled with the system of physical interest $A$; in this case one should
not consider $\gamma_t^A$ but $\gamma_t^A\otimes{\bf I}_B$ as the
effective time evolution acting not on a state $\rho_A$, but on the
effective initial state $\rho_{AB}$ of the compound system $A + B$.
It is the abstractness of such setting that makes the need of complete
positivity physically unpalatable in open quantum system
dynamics~\cite{pech}. 

More concretely, one may consider $A$ and $B$ as systems of the same
kind in (weak) interaction with a same environment and thus evolving
in time according to an approximate reduced dynamics of the form
$\gamma_t \otimes\gamma_t$.

Actually, there exist some experimental setups where this is the case and,
moreover, the compound system $A + B$ is initially prepared in a
maximally entangled state $\rho_{AB}$~\cite{ben2,ben3}.
Then, the question is whether, for 
$(\gamma_t\otimes\gamma_t)[\rho_{AB}]$ to remain positive, $\gamma_t$ need be
completely positive or not.

In Theorem 3 we shall prove 
that, in the case of $A$ and $B$
$d$-dimensional systems, this is indeed so: in order that $\gamma_t
\otimes \gamma_t$ be positivity preserving, $\gamma_t$ must be
completely positive. The argument in favour of the necessity of
complete positivity for semigroup dynamics of open quantum
systems results thus strengthened with respect to the argument based on
$\gamma_t \otimes {\bf I}_B$.

Complete positivity is formulated as a property of linear maps
$\Gamma$ on algebras of operators $X$ and by duality transferred to
the corresponding transformations $\gamma$ of quantum states, according to
\begin{equation}
\label{dual}
\Tr(\rho \Gamma[X]) = \Tr(\gamma[\rho]X).
\end{equation}
We shall consider states represented by density matrices $\rho$ and
restrict our attention to $d$-dimensional quantum systems 
so that the
operators $X$'s will be represented by $d \times d$ matrices as well as 
the $\rho$'s.

\noindent
{\bf Definition 1.}~\cite{take}\quad
{\it $\Gamma : M_d({\bf C})\rightarrow M_d({\bf C})$ is completely
positive if $\forall n \in {\bf N}$, the map $\Gamma \otimes {\bf I}_n$
preserves positivity on $M_d({\bf C})\otimes M_n({\bf C})$, where
$M_n({\bf C})$ is any $n \times n$ matrix algebra and ${\bf I}_n$ the
identity operation on it.}

In fact, one need not check all $n$ but just $n = d$ as stated by a
theorem of Choi~\cite{choi}

\noindent
{\bf Theorem 1.}\quad
{\it $\Gamma: M_d({\bf C})\rightarrow M_d({\bf C})$ is completely
positive if and only if $\Gamma \otimes {\bf I}_d$ 
is positivity preserving on $M_d({\bf C}) \otimes M_d({\bf C})$.}

\noindent
{\bf Remark 1.}\quad
If the map $\Gamma: M_d({\bf C})\rightarrow M_d({\bf C})$ is positivity
preserving, but not completely positive, then there is a positive
$X \in M_d({\bf C}) \otimes M_d({\bf C})$ such that 
$(\Gamma \otimes{\bf I}_d)[X]$ is not positive. 
If $\vert \psi \rangle$ is an eigenvector of 
$(\Gamma \otimes {\bf I}_d)[X]$
relative to a negative eigenvalue, via duality, we get
\begin{equation}
\label{horod}
\Tr\Bigl((\gamma \otimes {\bf I}_d)
[\vert \psi \rangle \langle \psi \vert]X\Bigr)
= \langle \psi \vert (\Gamma \otimes {\bf I}_d)[X]\vert \psi \rangle <
0. 
\end{equation}
Therefore the linear map $\gamma \otimes {\bf I}_d$, dual to $\Gamma
\otimes {\bf I}_d$, does not preserve the positivity of 
$\vert \psi \rangle\langle \psi \vert$.
Also, $\vert\psi\rangle$
must be entangled, for, if  
$\vert\psi\rangle=\vert \psi_a \rangle \langle \psi_a \vert
\otimes \vert \psi_b \rangle \langle \psi_b \vert$, then 
$\gamma [\vert \psi_a \rangle \langle \psi_a \vert] \otimes \vert
\psi_b \rangle \langle \psi_b \vert$ is positive.
\hfill$\square$

As stated in the introduction, we are interested in semigroups of
positive linear maps, $\{\gamma_t\}_{t\geqslant 0}$, on the states over
$M_d({\bf C})$. In particular, we shall be concerned with Hermiticity
and trace-preserving, continuous semigroups on density matrices
$\rho \in M_d({\bf C})$, 
\begin{eqnarray}
\label{semig1}
\gamma_t \circ \gamma_s &=& \gamma_{t + s} = \gamma_s \circ \gamma_t\ ,
\quad\forall s,t \geqslant 0\ ; \\
\label{semig2}
\Tr \gamma_t[\rho] &=& \Tr \rho,\; \gamma_t[\rho]^{\dagger} =
\gamma_t[\rho]\ ; \\
\label{semig3}
\lim_{t\rightarrow 0^+}\gamma_t[\rho] &=& \rho\ ,
\end{eqnarray}
the latter limit being understood in the trace-norm topology~\cite{gori}.

\noindent
{\bf Proposition 1.}~\cite{gori}\quad
{\it Any semigroup $\{\gamma_t\}$ satisfying~(\ref{semig1}-\ref{semig3}) 
is generated by the
equation:
\begin{equation}
\label{koss}
\partial_t \gamma_t[\rho] = -i \left[H, \gamma_t[\rho]\right] 
+ \sum_{a,b
= 1}^{d^2 -1} c_{ab}\left[F_a \gamma_t[\rho] F_b^{\dagger}
- \frac{1}{2} \{F_b^{\dagger} F_a, \gamma_t[\rho]\}\right]\ ,
\end{equation}
where $H = H^{\dagger}$, $\Tr H = 0$; $\Tr F_a^{\dagger} F_b =
\delta_{ab}$, $\Tr F_a =0$, $F_{d^2} = {\bf 1}_d/\sqrt{d}$ and
$C = [c_{ab}]$ is a $(d^2 -1) \times (d^2 - 1)$ self-adjoint matrix
depending solely on the choices of the traceless matrices 
$\{F_a\}_{a= 1}^{d^2 - 1}$.}
\smallskip

If one asks the $\gamma_t$ to be completely positive, that is dual to
completely positive $\Gamma_t: M_d({\bf C})\rightarrow M_d({\bf
C})$, then

\noindent
{\bf Theorem 2.}~\cite{gori}\quad
{\it The semigroup $\{\gamma_t \}_{t\geqslant 0}$ generated by~(\ref{koss})
consist of completely positive maps if and only if $C = [c_{ab}]$ is a
positive-definite $(d^2 -1) \times (d^2 - 1)$ matrix.}

\noindent
{\bf Remark 2.}\quad
If $C = [c_{ab}]$ is positive definite then it can be written $C =
A^{\dagger}A$, $c_{ab} = \sum_{r = 1}^{d^2 - 1} A^*_{ra} A_{rb}$, and
\begin{equation*}
\sum_{a,b = 1}^{d^2 - 1} c_{ab}\left[F_a \rho F_b^{\dagger} -
\frac{1}{2}\{F_b^{\dagger} F_a, \rho\}\right] = 
\sum_{r = 1}^{d^2 -
1}\left[V_r \rho V_r^{\dagger} - \frac{1}{2} \{V_r^{\dagger} V_r,
\rho\}\right]
\end{equation*}
takes the Lindblad form~\cite{lind} with 
$V_r = \sum_{a = 1}^{d^2 -
1} A_{ra}^* F_a$. Vice versa, given a generator in Lindblad's form, 
developing $V_r = \sum_{a = 1}^{d^2 - 1} v_{ra}F_a$ 
over a basis of traceless matrices $F_a$, one ends
up with a generator as in~(\ref{koss}) with $c_{ab} = \sum_{r =
1}^{d^2 - 1} V_{ra} V^*_{rb}$ making for a $(d^2 - 1) \times
(d^2 - 1)$ positive matrix $C = [c_{ab}]$.
\hfill$\square$

Given a semigroup $\{\gamma_t\}_{t\geqslant 0}$
satisfying~(\ref{semig1}-\ref{semig3}) and generated by~(\ref{koss}), the
justification why $\gamma_t$ should be completely positive and thus
the matrix $C = [c_{ab}]$ positive, is based on the fact that,
otherwise, $\gamma_t \otimes {\bf I}_d$ would fail to preserve the
positivity of entangled states on $M_d({\bf C}) \otimes M_d({\bf C})$
(see Remark 1).

However, while the first factor in 
$M_d({\bf C}) \otimes M_d({\bf C})$
refers to a concrete open quantum system evolving in time according
to~(\ref{koss}), because of the interaction with a certain
environment, the second factor represents a mere possibility of
entanglement with anything described by a $d$-dimensional system and
generically out of physical control. 

Instead, we argue that complete positivity is necessary to
avoid physical inconsistencies in compound systems consisting of two
$d$-dimensional systems that interact with a same environment, but not
among themselves, neither directly, nor indirectly, that is
through the environment itself.
In such a case, the two systems are expected to evolve according to
semigroups of linear maps 
$\gamma_t \otimes
\gamma_t$, $t\geqslant 0$, where $\gamma_t$ is the single
open system dynamics obtained when only one of them is present in the
environment. 

A necessary request for the physical consistency of such dynamics 
is that the $\gamma_t \otimes \gamma_t$'s preserve 
the positivity of all separable and entangled states of the compound system,
which now describe physically concrete and controllable settings.

\noindent
{\bf Theorem 3.}\quad
{\it If $\{\gamma_t\}_{t\geqslant 0}$ is a Hermiticity and trace preserving
continuous semigroup of linear maps over the states of $M_d({\bf C})$,
the semigroup $\{\gamma_t \otimes \gamma_t\}_{t\geqslant 0}$ of linear
maps over the states of $M_d({\bf C}) \otimes M_d({\bf C})$ is
positivity-preserving if and only if $\{\gamma_t\}_{t\geqslant 0}$ is
made of completely positive maps.}

The proof of Theorem 3 will consist of several steps. We need just
show the only if part; indeed, if $\gamma_t$  is completely positive,
$\gamma_t \otimes {\bf I}_d$ and ${\bf I}_d \otimes \gamma_t$ are both
positive and  such is the composite map
$\gamma_t \otimes \gamma_t = (\gamma_t \otimes
{\bf I}_d) \circ ({\bf I}_d \otimes \gamma_t)$.

\noindent
{\bf Remarks 3.}

\noindent
{\bf 1.}\quad
If the $\gamma_t$'s preserve the positivity of states of
$M_d({\bf C})$, $\gamma_t \otimes \gamma_t$ preserves the positivity
of separable states of $M_d({\bf C}) \otimes M_d({\bf C})$: this
follows by a straightforward adaptation of the argument in Remark 1.

\noindent
{\bf 2.}\quad
For generic positive linear maps $\gamma$ on the states of
$M_d({\bf C})$, it does not follow that, if $\gamma \otimes \gamma$
is positivity preserving, then $\gamma$ is completely positive.
A counter example is the transposition $\tau$ over $M_2({\bf C})$:
$\tau \otimes \tau$ is positivity-preserving, but $\tau$ is not
completely positive. 
We notice, however, that $\tau$ cannot be among the $\gamma_t$ of a
continuous semigroup over the states of $M_2({\bf C})$ since it is
not connected to the identity operation.

\noindent
{\bf 3.}\quad
There are experimental situations that are describable by
semigroups $\{\gamma_t \otimes \gamma_t\}_{t\geq0}$.
For instance neutral mesons may be imagined to suffer from
dissipative effects due to a noisy background determined by Planck's
scale physics. As decay products of spin $1$ resonances, these mesons 
are produced in maximally entangled states and, while
independently flying apart back to back, they arguably evolve
according to semigroups $\{\gamma_t \otimes
\gamma_t\}_{t\geq0}$~\cite{ben1,ben2,ben3}. 
In such a context, whether $\gamma_t \otimes \gamma_t$ is 
positivity-preserving is crucial for concrete physical consistency.
\hfill$\square$

\noindent
{\bf Lemma 1.}\quad
{\it If $\{\gamma_t\}_{t\geqslant 0}$ is a semigroup
satisfying~(\ref{semig1}-\ref{semig3}) 
and generated by~(\ref{koss}), the semigroup
$\{\gamma_t \otimes \gamma_t\}_{t\geqslant 0}$ consists of
positivity-preserving maps only if
\begin{equation}
\label{cond}
{\cal L}_{\phi, \psi} \equiv \langle \phi \vert (L\otimes {\bf I}_d +
{\bf I}_d \otimes L)[\vert 
\psi \rangle \langle \psi \vert] \vert \phi \rangle \geqslant 0
\end{equation}
for all orthogonal vector states $\vert \phi \rangle$, $\vert \psi
\rangle$ in ${\bf C}^d$, where $L$ is the generator on the right hand
side of~(\ref{koss}) and ${\bf I}_d$ is the identity operation on
$M_d({\bf C})$.}

{\bf Proof:}\quad
From the request of positivity preservation it follows that
${\cal G}_{\phi, \psi}(t):=$\hfill\break
$\langle \phi \vert (\gamma_t \otimes \gamma_t) 
[\vert \psi \rangle
\langle \psi \vert ] \vert \phi \rangle \geqslant 0$, 
for all  $\vert \phi\rangle$
and $\vert \psi \rangle\, \in {\bf C}^d \otimes {\bf C}^d$.
Choosing $\langle \phi \vert \psi \rangle = 0$, if 
${\rm d}{\cal G}_{\phi, \psi}(t)/{\rm d}t\vert_{t=0} < 0$, then
${\cal G}_{\phi, \psi}(t)\geqslant 0$ is violated in a neighborhood of $t=0$. 
Thus~(\ref{cond}) follows.\hfill $\blacksquare$

\noindent
{\bf Lemma 2.}\quad
{\it In the hypothesis of Lemma 1, let
$\{\vert j \rangle\}_{j = 1}^d$ be an orthonormal basis of ${\bf
C}^d$, and $\Phi$, $\Psi$ the $d \times d$
matrices $\Phi = [\varphi_{ij}]$, $\Psi = [\psi_{ij}]$
consisting of the coefficients of the expansion of $\vert \phi
\rangle$ and $\vert \psi \rangle$ with respect to the basis $\{\vert j
\rangle \otimes \vert k \rangle\}_{j,k = 1}^{d}$ of ${\bf C}^d \otimes
{\bf C}^d$.
Then
\begin{equation}
\label{dim1}
{\cal L}_{\phi, \psi} = 
\sum_{a,b = 1}^{d^2 - 1} c_{ab} \left[\Tr(\Psi
\Phi^{\dagger}F_a)\Tr(\Phi \Psi^{\dagger}F_b^{\dagger}) +  
\Tr((\Phi^{\dagger}\Psi)^T F_a)\Tr((\Psi^{\dagger} \Phi)^T
F_b^{\dagger}) \right]\ ,
\end{equation}
where $C=[c_{ab}]$ is the matrix of coefficient and
$F_a, F_b$ the traceless matrices appearing
in~(\ref{koss}), while $X^T$ denotes
transposition of $X$ with respect to the chosen basis.}

{\bf Proof:}\quad
Let $\vert \phi \rangle = \sum_{j,k = 1}^{d}\varphi_{jk} \vert j
\rangle \otimes \vert k \rangle$, 
$\vert \psi \rangle = \sum_{j,k =
1}^{d} \psi_{jk} \vert j \rangle \otimes \vert k \rangle$;
then, one calculates
\begin{equation}
\label{dim2}
\begin{split}
{\cal L}_{\phi, \psi} & =
\sum_{ij} \sum_{kl} \sum_{pr} (
\varphi^*_{ij} \varphi_{kl} \psi_{pj} \psi^*_{rl} + 
\varphi^*_{ji} \varphi_{lk} \psi_{jp} \psi^*_{lr}
) 
\langle
i \vert L[\vert p \rangle \langle r \vert] \vert k \rangle  
\\ 
& =
\sum_{ik} \sum_{pr} [
(\Psi \Phi^{\dagger})_{pi}(\Phi
\Psi^{\dagger})_{kr} +
(\Phi^{\dagger} \Psi)_{ip}(\Psi^{\dagger}
\Phi)_{rk} ] 
\langle i \vert L[\vert p \rangle \langle r \vert]
\vert k \rangle.
\end{split}
\end{equation}
The commutator and the anticommutator in
the generator $L[\cdot]$ drop from equation~(\ref{dim2}); this is
easily seen by noting that, given any $K \in M_d({\bf C})$,
$\langle i \vert (K \vert p \rangle\langle r \vert)\vert k \rangle =
K_{ip}\delta_{rk}$. In~(\ref{dim2}), we can further sum over
either $r = k$ or $i = p$; in either cases, as $\langle \phi
\vert \psi \rangle = 0$, we find $\Tr \Psi
\Phi^{\dagger} = (\Tr \Phi \Psi^{\dagger})^* = 0$ 
and the result follows. \hfill $\blacksquare$

\noindent
{\bf Lemma 3.}\quad
{\it In the hypothesis of Lemma 1, the matrix $C=[c_{ab}]$
in~(\ref{dim1}) must be positive definite.}

{\bf Proof:}\quad
With any $\vec{w} = \{ w_a  \}_{a = 1}^{d^2 - 1}\in{\bf
C}^{d^2 -1}$, we consider $W = \frac{1}{2}
\sum_{a = 1}^{d^2 - 1} w_a^* F_a$, which is a traceless $d\times d$ matrix.
If matrices $\Psi$ and $\Phi$ exist such that $\Phi \Psi^{\dagger} = W$
and $\Psi^{\dagger} \Phi = W^T$, then, from Lemmas
1 and 2 and the orthogonality of the matrices $F_a$ (compare
Theorem 2) it follows
\begin{equation}
\label{dim4}
{\cal L}_{\phi, \psi} = \sum_{a,b = 1}^{d^2 - 1} c_{ab} w^*_a w_b
\geqslant 0\ ,
\end{equation}
whence the positivity of $C = [c_{ab}]$ and the proof of Theorem 3.
Any matrix $W$ and its transposed with respect to the given basis, $W^T$,
have the same elementary divisors; therefore, they are similar to the same
canonical Jordan form and thus similar to each other~\cite{gelf}.
Let $\Phi$ such that $\Phi^{-1} W \Phi = W^T$, that is we take as
vector $\vert \phi \rangle \in {\bf C}^d \otimes {\bf C}^d$ the one
whose components $\phi_{ij}$ are the elements of the similarity matrix
transforming the given $W$ into its transposed $W^T$.
It then follows that $\Psi^{\dagger} = \Phi^{-1} W$ and moreover
$\Psi^{\dagger} \Phi = \Phi^{-1} W \Phi = W^T$, which is what we
need. \hfill $\blacksquare$ 

\noindent
{\bf Remarks 4.}

\noindent
{\bf 1.}\quad
In the proof of Theorem 2 in~\cite{gori}, 
the maximally entangled state
$\vert \phi_+ \rangle = \frac{1}{d} \sum_{i = 1}^{d} \vert i \rangle
\otimes \vert i \rangle$ plays a crucial role; however, 
it concerns a generator of the form $L \otimes {\bf
I}_d$ instead of $L \otimes {\bf I}_d + {\bf I}_d \otimes L$.
In such a case,~(\ref{dim1}) reads 
\begin{equation*}
{\cal L}_{\phi, \psi} = \sum_{a,b = 1}^{d^2 - 1} c_{ab} \Tr(\Psi
\Phi^{\dagger} F_a)[\Tr(\Psi \Phi^{\dagger} F_b)]^*.
\end{equation*}
Choosing $\Phi = \Phi^{\dagger} = {\bf 1}_d/d$ given by the
components of $\vert \phi_+ \rangle$ and $W = \Psi^\dagger = d \sum_{k =
1}^{d^2 - 1} w^*_kF_k$, the result of Theorem 2 in~\cite{gori} 
immediately follows from our argument, for 
\begin{equation*}
{\cal L}_{\phi, \psi} = \sum_{a,b = 1}^{d^2 - 1} c_{ab} w_a^*
w_b \geqslant 0. 
\end{equation*}

\noindent
{\bf 2.}\quad
The choice of $\Phi = {\bf 1}_d/d$ in the previous
Remark is fixed for all traceless matrices $W$ and it is $\Psi^\dagger$ 
which is
chosen to be $W$.
This argument, however, does not
work with generic ${\cal L}_{\phi, \psi}$ as
in~(\ref{cond}), for, in general, $W^T \neq W$.
Nevertheless, when the $F_a$'s are self-adjoint and $C = [c_{ab}]$ 
symmetric, the choice $\Phi = {\bf 1}_d/d$ suffices for
proving Theorem 3.
In this case, positivity of $C = [c_{ab}]$ is checked against real
vectors $\vec{w} \in {\bf R}^{d^2 - 1}$, so that one can restrict
to self-adjoint $W = W^\dagger$ and choose a basis $\{\vert i \rangle\} 
\in {\bf C}^d$ such that $W$ is diagonal; then $(\Psi^{\dagger})^T =W^T=W$.

\noindent
{\bf 3.}\quad
When $d = 2$, the maximally entangled Bell state $\vert \phi \rangle =
(\vert 1 \rangle \otimes \vert 2 \rangle - \vert 2 \rangle
\otimes \vert 1 \rangle)/\sqrt{2}$ plays for the generator $L\otimes {\bf
I}_d+{\bf I}_d\otimes L$ the same role played by the symmetric state
$|\phi_+\rangle$ for the generator $L\otimes {\bf I}_d$ in 
Remark 4.1~\cite{gori}.
Namely, given any traceless matrix 
$W = \left(
\begin{array}{cc} \alpha & \beta \\
\gamma & -\alpha \end{array}
\right)$, we can choose 
$\Phi =
\frac{1}{\sqrt{2}} \left(
\begin{array}{cc} 0 & 1 \\
-1 & 0 \end{array}
\right)$ and
$\Psi^{\dagger} = \Phi^{-1} W = \sqrt{2} \left(
\begin{array}{cc} -\gamma & \alpha \\
\alpha & \beta \end{array}
\right)$.
It turns out that
$\Psi^{\dagger} \Phi = 
\left(
\begin{array}{cc} -\alpha & -\gamma \\
-\beta & \alpha \end{array}
\right) = -W^T$~\cite{halm} and the minus sign is not felt by the 
expressions in~(\ref{dim1}).
\hfill$\square$

%%%%%%%%%%%%%%%%%%%%%%%%%%%%%%%%%%%%%%%%%%%%%%%%%%%%%%%%%%%%%%%%%%%%%%%%%%%%%%

\end{document}